\documentclass[a4paper,11pt]{article} 
\usepackage{graphicx}
\usepackage{subfigure}
\usepackage{helvet}
\usepackage[utf8]{inputenc}
\usepackage{mathtools}
\usepackage[T1]{fontenc}
\usepackage{geometry}
\usepackage[english]{babel}
\usepackage{latexsym,bm}
\usepackage{placeins}
\usepackage{enumitem}
\usepackage[sort&compress]{natbib}
\usepackage{cite}
\usepackage{xparse}
\usepackage{color}
\usepackage{extarrows}
\usepackage{chngcntr}
\usepackage{mathrsfs}
\usepackage[font={small}]{caption}
\usepackage{tikz}
\usepackage{setspace}
\usepackage{amsmath,scalefnt}
\usepackage{amssymb}
\usepackage{ctable}
\usepackage{array}
\usepackage{theorem}
\usepackage{algorithm}
\usepackage{algpseudocode}

\usetikzlibrary{arrows}
\usetikzlibrary{shapes.geometric}
\usetikzlibrary{shapes.multipart}
\usetikzlibrary{positioning}
\usetikzlibrary{trees}

\newcolumntype{M}[1]{>{\centering\arraybackslash}m{#1}}
\newcolumntype{N}{@{}m{0pt}@{}}

\newlength{\MMtextNodeWidth}

\newenvironment{vartalgorithm}[1]
  {\algorithm[t]\renewcommand{\thealgorithm}{#1}}
  {\endalgorithm}

\newenvironment{Tabular}[2][1]
  {\def\arraystretch{#1}\tabular{#2}}
  {\endtabular}

\newcommand{\listequationnumber}{\refstepcounter{equation}(\theequation)}
\newcommand{\listequation}[1]{$\displaystyle #1$\hfill\listequationnumber}
\newcommand*\myat{{\fontfamily{ptm}\selectfont @}}

\newcommand{\N}{{\rm N}}
\newcommand{\D}{{\rm d}}
\newcommand{\M}{\mathcal{F}}

\newcommand{\yobs}{Y}
\newcommand{\ymis}{W}
\newcommand{\tymis}{\tilde{W}}
\newcommand{\cutt}{(t)}
\newcommand{\ini}{(0)}
\newcommand{\nextt}{(t+1)}

\newcommand{\T}{{\rm T}}
\newcommand{\logr}{{\rm log}}
\newcommand{\order}{{\rm th}}
\newcommand{\tr}{{\rm trace}}
\newcommand{\ivd}{IvD }
\newcommand{\bn}{BN }

\setdescription{font=\normalfont}

\DeclareDocumentEnvironment{steps}%
{O{Step}}
{\begin{enumerate}[label=#1 \arabic*,leftmargin=39pt]}%
{\end{enumerate}}

\makeatletter
\def\step{%
  \@ifnextchar[\@step{\@noitemargtrue\@step[\@itemlabel]}} 
\def\@step[#1]{\item[#1:]}
\makeatother

\makeatletter
\def\old@comma{,}
\catcode`\,=13
\def,{%
  \ifmmode%
   \old@comma\discretionary{}{}{}%
  \else%
    \old@comma%
  \fi%
}
\makeatother

\title{\textbf{\LARGE{A Corrected and More Efficient Suite of MCMC Samplers for the Multinomal Probit Model}}}
\author{
Xiyun~Jiao and David~A.~van~Dyk
\thanks{\protect\doublespacing
{Xiyun~Jiao is a postgraduate student in Statistics, Department of Mathematics, Imperial College London, SW7 2AZ (Email: {\bf x.jiao12\myat imperial.ac.uk}); Professor David~A.~van~Dyk holds a Chair in Statistics, Department of Mathematics, Imperial College London (Email: {\bf d.van-dyk\myat imperial.ac.uk}).}
}\\
Statistics Section, Department of Mathematics, Imperial College London
}

\date{}
\begin{document}
\maketitle 

\begin{abstract}
The multinomial probit (MNP) model is a useful tool for describing discrete-choice data and there are a variety of methods for fitting the model. Among them, the algorithms provided by \citet{imai:vand:05}, based on Marginal Data Augmentation, are widely used, because they are efficient in terms of convergence and allow the possibly improper prior distribution to be specified directly on identifiable parameters. \citet{burg:nord:12} modify a model and algorithm of \citet{imai:vand:05} to avoid an arbitrary choice that is often made to establish identifiability. There is an error in the algorithms of \citet{imai:vand:05}, however, which affects both their algorithms and that of \citet{burg:nord:12}. This error can alter the stationary distribution and the resulting fitted parameters as well as the efficiency of these algorithms. We propose a correction and use both a simulation study and a real-data analysis to illustrate the difference between the original and corrected algorithms, both in terms of their estimated posterior distributions and their convergence properties. In some cases, the effect on the stationary distribution can be substantial.
\end{abstract}



\clearpage
\section{Introduction}
\label{sec:intro}
The multinomial probit (MNP) model is widely used for describing discrete-choice data in social sciences and transportation studies. It is often preferred over the multinomial logit model because it does not assume independence of irrelevant alternatives; see, e.g., \citet{haus:wise:78} for details. Moreover, the MNP model has a strong connection with the multiperiod probit model, for which binary choices are observed over multiple time periods with correlated errors \citep{mcc:rossi:94}. 

The use of the MNP model was once restricted, because methods, like maximum likelihood estimates or simulated moments \citep{mcf:89}, require evaluating high-dimensional normal integrals, which are typically intractable. More recently, advances in Bayesian simulations have boosted the development of Markov chain Monte Carlo (MCMC) algorithms for fitting the MNP model (e.g., \citet{mcc:rossi:94}, \citet{nob:98}, \citet{mcc:etal:00}, \citet{imai:vand:05}, and \citet{burg:nord:12}). These algorithms avoided evaluating multidimensional integrals, provided reliable model fitting, and thus revitalized the use of the MNP model in practice.

Current MCMC algorithms specify a set of latent Gaussian variables as augmented data, whose relative magnitudes determine the choices. Since the augmented model is not identifiable given the observations, a proper prior distribution is required
to ensure that the posterior distribution is proper. \citet{mcc:rossi:94} advocate a Gibbs sampler which was the first feasible Bayesian approach to fitting the MNP model. In their specification, however, the prior distribution for the identifiable parameters is only determined as a byproduct \citep[][henceforth IvD]{imai:vand:05}. An improvement of \citet{mcc:rossi:94} is the ``hybrid Markov chain'' introduced by \citet{nob:98}, which adds a Metropolis step to sample the unidentifiable parameters. \citet{mcc:etal:00} propose another modification of \citet{mcc:rossi:94} which specifies a prior distribution directly on the identifiable parameters. \ivd review these MCMC algorithms, compare their computational performance, and find that, first, \citet{nob:98} can be less sensitive to starting values than \citet{mcc:rossi:94}; second, although Nobile's method significantly improves the convergence of the overall chain, the gain seems to be primary for the unidentifiable parameter with only slight gain for the identifiable ones; and third, although \citet{mcc:etal:00} solve the problem of prior specification, their algorithm can be less efficient in terms of convergence than either \citet{mcc:rossi:94} or \citet{nob:98} (This final point was also noted by \citet{mcc:etal:00} and \citet{nob:00}.) Moreover, \ivd point out an error in Nobile's derivation which can alter its stationary distribution. Ironically, as we shall see, the algorithms of \ivd also contain an error.  

\ivd develop new samplers based on the Marginal Data Augmentation (MDA) algorithm \citep{meng:vand:99}. The new algorithms are easy to implement because they only include draws from standard distributions. \ivd demonstrate that first, their methods are at least as quick as the fastest methods in terms of convergence, and second, the model is specified in terms of possibly improper prior distributions that are set directly on the identifiable parameters, making the priors relatively easy to interpret. Because of their apparent advantages, IvD's algorithms have been widely used in practice to fit MNP models; see, e.g., \citet{berr:cald:12}, \citet{burg:nord:12}, \citet{chau:14}, \citet{hori:etal:07}, \citet{hrus:07}, \citet{lu:etal:12}, \citet{queralt:12}, \citet{sin:whit:13}, \citet{vinc:etal:13}, \citet{zhang:etal:08}, etc. This success has been aided by a popular R package (\emph{MNP}, \citet{imai:vand:05jss}).

Unfortunately, there are two errors in IvD's algorithms; both occur when sampling the variance-covariance matrix. First, \ivd reparameterize the variables to facilitate the sampling of the variance-covariance matrix, and they make a mistake when transforming to the original parameterization. Second, when updating the variance-covariance matrix, a constraint on the matrix is overlooked. These errors can alter the stationary distribution and hence the fitted values and standard errors of the model parameters. They also can affect the efficiency of convergence. 

\citet[][henceforth BN]{burg:nord:12} modify the model of \ivd by changing the manner in which unidentifiability in the scale is addressed. In particular, they fix the trace of the variance-covariance matrix while IvD, like previous authors, fix the first diagonal element. BN's algorithm for sampling from the posterior distribution builds upon Algorithm~1 of IvD. Thus the two errors made by \ivd also affect BN's algorithm. \bn even make another mistake when updating the regression coefficient parameter, $\beta$. In this paper, we explain how to correct the errors in algorithms of both \ivd and BN, and use both a simulation study and a real-data analysis to illustrate the difference between the original and the corrected algorithms in terms of their estimated posterior distributions and convergence properties. The corrections we propose will be implemented in the \emph{MNP} R package.

The remainder of this paper is organized into five sections. We introduce the MNP model in Section~\ref{sec:model}. In Section~\ref{sec:algo}, we present the original algorithms in \ivd and BN, point out their errors, and provide the corrected algorithms. We also include a short review of MDA, which is used by all the algorithms we consider. In Sections~\ref{sec:simu} and~\ref{sec:real}, we use a simulation study and a real-data example, respectively, to illustrate the difference between the original and corrected algorithms. Conclusion and final remarks appear in Section~\ref{sec:con}.

\section{Multinomial Probit Model}
\label{sec:model}
We consider a $(p+1)$-class multinomial model. Each observation is a binary $(p+1)$-vector, $d_i=(d_{i1}, \dots, d_{i,(p+1)})$. We model $d_i$ by conditioning on a latent multivariate normal variable, $U_i=\left(U_{i1},\dots,U_{i,(p+1)}\right)$; $d_{ij}$ is one if $U_{ij}$ is larger than all the other components of $U_i$. Specifically, 
\begin{equation}
U_i\sim\N_{p+1}\left({X_i^0}\beta, \Sigma^0\right)\mbox{ and }
d_{ij}=\left\{
  \begin{array}{l l}
   1 & \quad \text{if $U_{ij}={\rm max}\{U_{i1},\dots,U_{i,(p+1)}\}$}\\
   0 & \quad \text{otherwise}
  \end{array}, \right.\ \mbox{for}\ i=1,\dots,n,
\label{eq:model1}
\end{equation}
where $X_i^0$ is a $((p+1)\times q)$ matrix of known covariates, $\beta$ is a $q$-vector of unknown parameters, and $\Sigma^0$ is a $((p+1)\times (p+1))$ unknown variance-covariance matrix. 

Model~(\ref{eq:model1}) is unidentifiable because shifting $U_i$ by any constant or rescaling $U_i$ by any positive constant, does not alter the distribution of $d_i$. To avoid this, \ivd and \bn both follow \citet{mcc:rossi:94}, by expressing each $U_{ij}$ relative to a base category (e.g., $U_{i,(p+1)}$), and obtain the new latent variable, $W_i=(W_{i1},\dots,W_{ip})$, where $W_{ij}=U_{ij}-U_{i,(p+1)}$. The distribution of $W_i$ is still multivariate normal, that is, 
\begin{equation}
W_{i}\sim\N_{p}(X_i \beta, \Sigma),
\label{eq:w}
\end{equation}
 where $X_i=PX_i^0$ and $\Sigma=P\Sigma^0 P^\T$ with $P=[I_p,-J]$, with $I_p$ a $(p\times p)$ identity matrix and $J$ a column $p$-vector of ones. For simplicity, we collapse $d_i$ into $Y_i$, which is an integer in $\{0,\dots,p\}$, defined as   
\begin{equation}
Y_{i}=\left\{
  \begin{array}{l l}
   0 & \quad \text{if max\{$W_{i1},\dots,W_{ip}$\}$<0$}\\
   k & \quad \text{if $W_{ik}={\rm max}\{0,W_{i1},\dots,W_{ip}\}$}
  \end{array}, \right.\ \mbox{for}\ i=1,\dots,n.
\label{eq:fmodel}
\end{equation}

To ensure identifiability, we must also set the scale. \ivd adopt the standard solution of \citet{mcc:rossi:94}; they set the first diagonal element of $\Sigma$ to one, i.e., $\sigma^2_{11}=1$. \bn propose a different solution; they fix the trace of the variance-covariance matrix, i.e., $\tr(\Sigma)=p$. They argue that the trace restriction is a better choice from three reasons. First, the trace restriction makes it possible to specify a symmetric prior for $\Sigma$ that is invariant to permutations of rows and columns. Second, when using the variance-element restriction (as in IvD), the estimated predicted choice probabilities under the posterior distribution can vary largely with the choice of the category corresponding to the unit variance. The trace-restricted fits tend to be intermediate among the results of the $p$ possible variance-element restricted fits. Third, the trace restriction yields marginal posterior distributions that are easier to interpret.

To overcome difficulties stemming from the constraint, $\sigma^2_{11}=1$, on the variance-covariance matrix, motivated by \citet{mcc:rossi:94}, \ivd set $\tilde{W}_i=\alpha W_i$, for $i=1,\dots,n$, where $\alpha>0$. Then $\tilde{W}_i\sim \N_{p}(X_i \tilde{\beta}, \tilde{\Sigma})$, where $\tilde{\beta}=\alpha \beta$ and $\tilde{\Sigma}=\alpha^2\Sigma$. Because $\tilde{\Sigma}$ can be any positive-definite matrix, \ivd specify an $\mbox{inverse-Wishart}$ prior distribution, $\tilde{\Sigma}\sim \mbox{Inv-Wishart}(\nu,\tilde{S})$. After transforming to $\alpha^2=\tilde{\sigma}^2_{11}$ and $\Sigma=\tilde{\Sigma}/\tilde{\sigma}^2_{11}$, the implied prior distribution on $(\alpha^2,\Sigma)$ is 
\begin{equation}
\alpha^2|\Sigma\sim \alpha_0^2\tr(S\Sigma^{-1})/\chi^2_{\nu p},\mbox{ and }
p(\Sigma)\propto {|\Sigma|}^{-(\nu+p+1)/2}{[\tr(S\Sigma^{-1})]}^{-\nu p/2}I\{\sigma^2_{11}=1\},
\label{eq:prior}
\end{equation}
where $S=\tilde{S}/\alpha_0^2$ and the first diagonal element of $S$ is one; $I$ is an indicator function which equals one when the condition in the brackets is satisfied, and zero otherwise. They also specify a normal prior distribution for $\beta$, $\beta\sim \N_q(\beta_0,A)$. For simplicity, we set $\beta_0=0$. \bn adopt the same strategy for setting their prior distribution in the context of the constraint, $\tr(\Sigma)=p$. In particular, their implied prior distribution for $(\alpha^2,\Sigma)$ is almost the same as \ivd except that
\begin{equation}
p(\Sigma)\propto {|\Sigma|}^{-(\nu+p+1)/2}{[\tr(S\Sigma^{-1})]}^{-\nu p/2}I\{\tr(\Sigma)=p\},
\label{eq:priors}
\end{equation}
where $\tr(S)=p$. They use the same prior distribution as \ivd for $\beta$, i.e., $\beta\sim \N_q(0,A)$. As \ivd state, this choice of prior distribution allows both informative and diffuse priors for unknown parameters while maintaining simplicity and efficiency of the algorithms.

\section{MDA Algorithms for Fitting MNP Models}
\label{sec:algo}

\subsection{Marginal Data Augmentation}
\label{sec:mda}
The algorithms of \ivd and \bn are all based on the method of MDA. To describe and correct the errors in these algorithms, we briefly review MDA. First, denoting $(\beta,\Sigma)$ by $\theta$, the Data Augmentation (DA) algorithm \citep{tann:wong:87} is designed to sample from the posterior distribution, $p(\theta, W|Y)$, by updating from $p(\ymis|\theta,\yobs)$ and $p(\theta|\ymis,\yobs)$ iteratively. In this section, we regard $\yobs$, $\theta$ and $W$ as generic observed data, unknown parameter of interest, and latent variables, respectively. 

Although easy to implement, the DA algorithm can be slow to converge. The MDA algorithm \citep{meng:vand:99} improves the convergence rate of a standard DA algorithm by expanding its state space. Specifically, MDA introduces a working parameter, $\alpha$, into the augmented-data model $p(\ymis,\yobs|\theta)$; $\alpha$ is not identifiable under the observed-data model $p(\yobs|\theta)$. An MCMC sampler is run on the expanded model $p(\tymis,\yobs|\theta,\alpha)$, which is designed to maintain $p(\yobs|\theta)$ as its marginal distribution, that is, 
\begin{equation}
\int p(\tymis,\yobs|\theta,\alpha)\D\tymis=p(\yobs|\theta).
\label{eq:expand}
\end{equation}
A general method for introducing $\alpha$ into an augmented-data model is to use a one-to-one mapping,
\begin{equation}
\tymis=\M_{\alpha}(\ymis),\mbox{ for any given }\alpha,
\label{eq:functiona}
\end{equation}
which is differentiable when $\ymis$ is continuous. For each $\M$, there typically exists a value $\alpha_0$ such that $\M_{\alpha_0}$ is an identity function, $\M_{\alpha_0}(\ymis)=\ymis$. With this construction, the MDA algorithm proceeds by iterating
\begin{description}
\item[Step 1:] \listequation{(\tymis^{\nextt},\alpha^\star)\sim p(\tymis,\alpha|\theta^{\cutt},\yobs),}
\label{eq:sampler1}
\item[Step 2:] $(\theta^{\nextt},\alpha^{\nextt})\sim p(\theta,\alpha|\tymis^{\nextt},\yobs)$.
\end{description}
Note that in the sampler in~(\ref{eq:sampler1}), $\alpha$ is sampled in both steps and its first update is not part of the final output. We define such updates as \emph{intermediate quantities} and indicate them with superscript ``$\star$''. The sampler in~(\ref{eq:sampler1}) is a \emph{collapsed DA sampler} \citep{liu:wong:kong:94}, since its two steps can be considered as sampling $\tymis$ and $\theta$ with $\alpha$ integrated out. In this regard, the sampler in~(\ref{eq:sampler1}) is equivalent to a standard DA sampler constructed for the conditional distributions of $p(\tymis,\theta|\yobs)$. Thus the marginal Markov chain, $\{\theta^{\cutt}, t=0,1,\dots\}$, produced by the sampler in~(\ref{eq:sampler1}) is reversible with $p(\theta|\yobs)$ as its stationary distribution. Collapsing $\alpha$ out increases the (expected) variance of the conditional distributions sampled in~(\ref{eq:sampler1}). This enables bigger jumps and faster convergence, see \citet{meng:vand:99} and \citet{vand:meng:01art} for more details. 

\begin{vartalgorithm}{1.1}
\caption{}
\label{alg11}
\begin{algorithmic}
\footnotesize{\State {\bf Step~0:} Initialize parameters $t=0$, $\beta^{\ini}$, $\alpha^{\ini}$, $\Sigma^{\ini}$ and $W^{\ini}$.
\smallskip
\While {$t<T$}
\smallskip
\State {\bf Step~1:} Update $\left({(\alpha^2)}^{\star},\tilde{W}^\star\right)$ via $p(\alpha^2,\tilde{W}|Y,\beta^{\cutt}, \Sigma^{\cutt})$ by

\State (a) sampling ${(\alpha^2)}^{\star}$ from $p(\alpha^2|\Sigma^{\cutt})$: ${(\alpha^2)}^{\star}\sim \alpha_0^2\tr\left(S{\Sigma^{\cutt}}^{-1}\right)/\chi^2_{\nu p}$;

\State (b) sampling $\tilde{W}^\star$ from $p(\tilde{W}^\star|Y,{(\alpha^2)}^{\star},\beta^{\cutt},\Sigma^{\cutt})$: 
\For{$i:=1,\dots,n$}
\For{$k:=1,\dots,p$}

\State sampling $W_{ik}^\star$ from $p(W_{ik}|Y_i,W_{i,-k}^\star,\beta^{\cutt}, \Sigma^{\cutt})$: $W_{ik}^\star\sim {\rm TN}(\mu_{ik},\tau_{ik}^2)$, see Appendix~\ref{app:c} for details;
\EndFor

\State Set $\tilde{W}_i^\star=\alpha^{\star}{W}_i^\star$. 
\EndFor

\smallskip
\State {\bf Step~2:} Update $\left({(\alpha^2)}^{\star},\beta^{\nextt}\right)$ via $p(\alpha^2,\beta|Y,\tilde{W}^\star, \Sigma^{\cutt})$ by 

(a) sampling ${(\alpha^2)}^{\star}$ from $p(\alpha^2|Y,\tilde{W}^\star, \Sigma^{\cutt})$:
$${(\alpha^2)}^{\star}\sim
\frac{\sum_{i=1}^n{(\tilde{W}_i^\star-X_i\hat{\beta})}^\T{\Sigma^{\cutt}}^{-1}(\tilde{W}_i^\star-X_i\hat{\beta})
+\hat{\beta}^\T A^{-1}\hat{\beta}+\tr\left(\tilde{S}{\Sigma^{\cutt }}^{-1}\right)}{\chi^2_{(n+\nu) p}},$$
\State where $\hat{\beta}={\left(\sum_{i=1}^n X_i^\T {\Sigma^{\cutt }}^{-1}X_i +A^{-1}\right)}^{-1}
\left(\sum_{i=1}^n X_i^\T {\Sigma^{\cutt }}^{-1}\tilde{W}_i^\star\right)$;

(b) sampling $\tilde{\beta}^\star$ from $p(\tilde{\beta}|Y,\tilde{W}^\star, {(\alpha^2)}^\star,\Sigma^{\cutt})$:
$$\tilde{\beta}^\star\sim \N_q\left[\hat{\beta},
{(\alpha^2)}^{\star}{\left(\sum_{i=1}^n X_i^\T {\Sigma^{\cutt }}^{-1}X_i +A^{-1}\right)}^{-1}\right],$$
\State and setting $\beta^{\nextt}=\tilde{\beta}^\star/\alpha^{\star}$.

\smallskip
\State {\bf Step~3:} Update $\left({(\alpha^2)}^{\nextt},\Sigma^{\nextt}\right)$ via $p(\alpha^2,\Sigma|Y,\tilde{W}^\star, \beta^{\nextt})$ by 

(a) sampling $\tilde{\Sigma}^\star$ from $p(\tilde{\Sigma}|Y,Z,\beta^{\nextt})$: 
$$\tilde{\Sigma}^\star\sim \mbox{Inv-Wishart}\left(n+\nu,\sum_{i=1}^{n}Z_i Z_i^\T\right),$$
\State where $Z_i=\tilde{W}_i^\star-\alpha^\star X_i \beta^{\nextt}$; 

(b) setting ${\alpha}^{\nextt}=\tilde{\sigma}_{11}^\star$, $\Sigma^{\nextt}=\tilde{\Sigma}^\star/{\left(\alpha^{\nextt}\right)}^2$, and $W^{\nextt}=\tilde{W}^\star/\alpha^{\nextt}$.

\medskip
\State \Return $\beta^{\nextt}$, $\Sigma^{\nextt}$ and $W^{\nextt}$
\State $t+1 \gets t$
\EndWhile}
\end{algorithmic}
\end{vartalgorithm}

\begin{figure}[t]
\begin{center}
  \includegraphics[height=3.8in]{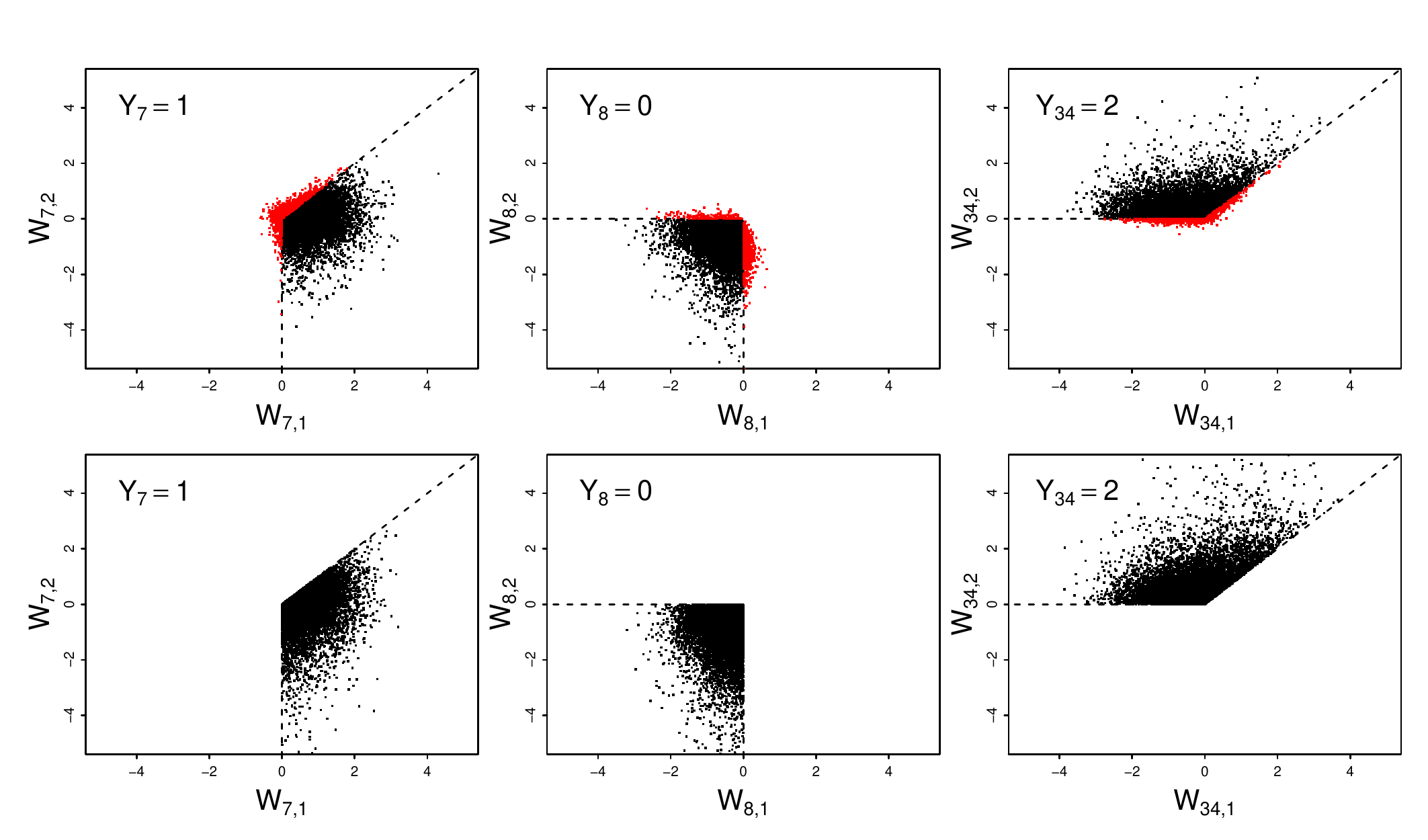}
\end{center}
    \caption{The posterior samples of $W_{7}$, $W_{8}$, and $W_{34}$ obtained with Algorithms~1.2 and~1.3 appear in the first and second rows, respectively. The samples from Algorithm~1.2 not adhering to the constraint~(\ref{eq:cons}) are plotted in red.}
\label{fig:1}
\end{figure}

\begin{figure}[t]
\begin{center}
\includegraphics[height=3.9in]{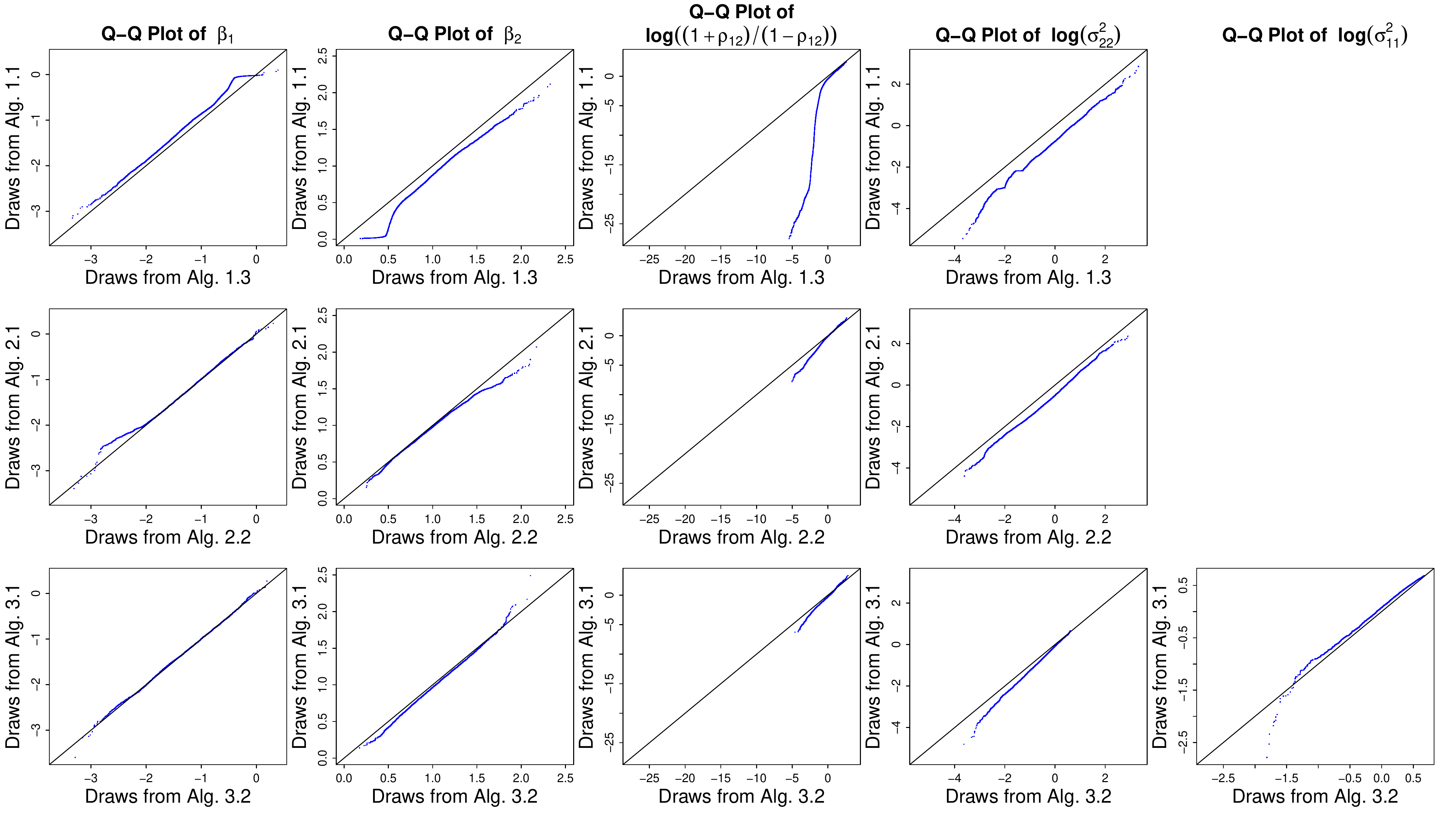}
\end{center}
\caption{Quantile-quantile plots for comparing posterior draws from different algorithms in the simulation study. The columns correspond to five parameters, i.e., $\beta_1$, $\beta_2$, $\logr\left(\frac{1+\rho_{12}}{1-\rho_{12}}\right)$, $\logr(\sigma^2_{22})$, and $\logr(\sigma^2_{11})$. The first row compares draws from Algorithms~\ref{alg11} and~1.3, the second row compares draws from Algorithms~\ref{alg21} and~2.2, and the last row compares draws from Algorithms~\ref{alg31} and~3.2.}
\label{fig:9}
\end{figure}

\begin{figure}[t]
\begin{center}
\includegraphics[height=4.5in]{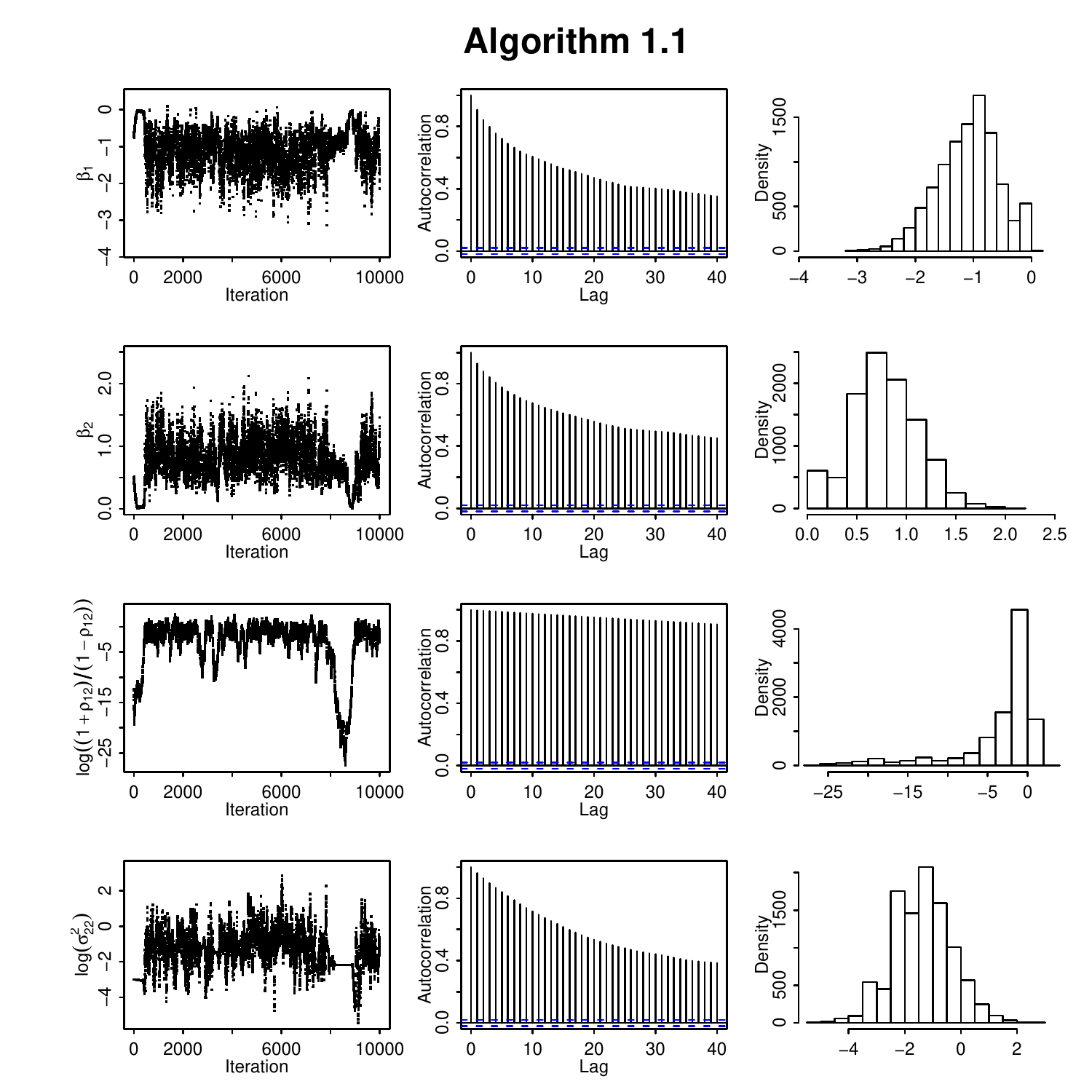}
\end{center}
\caption{The sampling results of Algorithm~\ref{alg11} for the simulation study. The columns correspond to trace plots, autocorrelation plots, and histograms. The rows correspond to four parameters: $\beta_1$, $\beta_2$, $\logr\left(\frac{1+\rho_{12}}{1-\rho_{12}}\right)$, and $\logr(\sigma^2_{22})$.}
\label{fig:2}
\end{figure}

\begin{figure}[t]
\begin{center}
\includegraphics[height=4.5in]{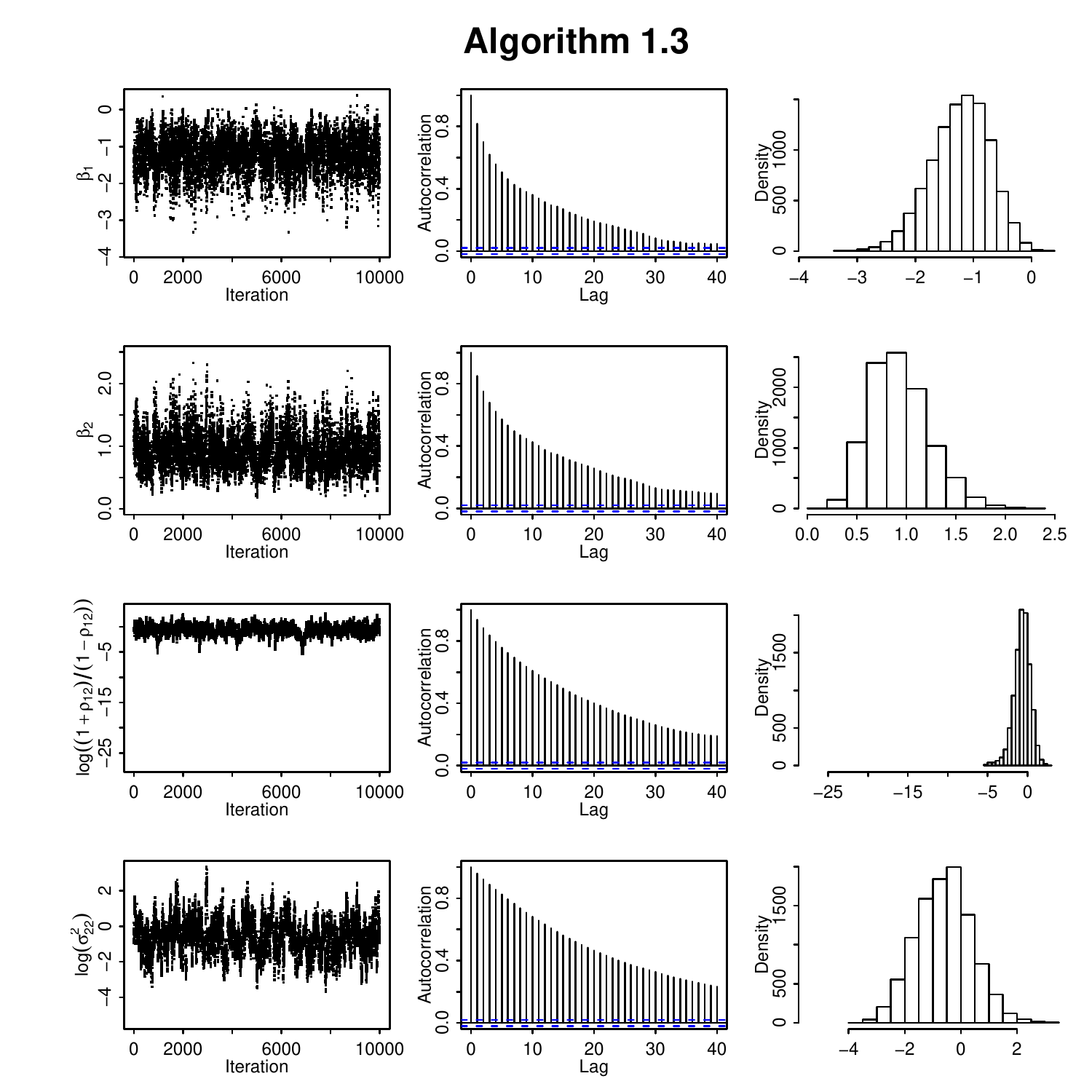}
\end{center}
\caption{The sampling results of Algorithm~1.3 for the simulation study. The columns correspond to trace plots, autocorrelation plots, and histograms. The rows correspond to four parameters: $\beta_1$, $\beta_2$, $\logr\left(\frac{1+\rho_{12}}{1-\rho_{12}}\right)$, and $\logr(\sigma^2_{22})$.}
\label{fig:4}
\end{figure}

\renewcommand{\arraystretch}{0.3}
\begin{table}[h]
\begin{centering}
\begin{Tabular}[2]{cccccccc}
\specialrule{.1em}{.05em}{.05em}
& {\bf Alg.~1.1} & {\bf Alg.~1.2} & {\bf Alg.~1.3} & {\bf Alg.~2.1} & {\bf Alg.~2.2} & {\bf Alg.~3.1} & {\bf Alg.~3.2}\\

\specialrule{.1em}{.05em}{.05em}

$\beta_1$ & 0.693 & 1.592 & 1.610 & 1.575 & 2.111 & 1.935 & 2.573\\

$\beta_2$ & 0.443 & 1.584 & 1.305 & 1.384 & 1.342 & 2.492 & 3.413\\

$\logr\left(\sigma_{11}^2\right)$ & - & - & - & - & - & 1.445 & 1.351\\

$\logr\left(\frac{1+\rho_{12}}{1-\rho_{12}}\right)$ & 0.060 & 0.625 & 0.865 & 0.782 & 1.070 & 0.768 & 1.163\\

$\logr\left(\sigma_{22}^2\right)$ & 0.506 & 1.334 & 0.703 & 1.474 & 0.823 & 1.261 & 1.216\\

\specialrule{.1em}{.05em}{.05em}
\end{Tabular}
\caption{Effective sample size per second for each of five parameters, i.e., $\beta_1$, $\beta_2$, $\logr\left(\sigma_{11}^2\right)$, $\logr\left(\frac{1+\rho_{12}}{1-\rho_{12}}\right)$, and $\logr\left(\sigma_{22}^2\right)$ in the simulation study, as obtained with Algorithms~\ref{alg11}--1.3, Algorithms~\ref{alg21}--2.2, and Algorithms~\ref{alg31}--3.2 respectively.}
\label{table:ess}
\end{centering}
\end{table}

\begin{figure}[t]
\begin{center}
\includegraphics[height=4.4in]{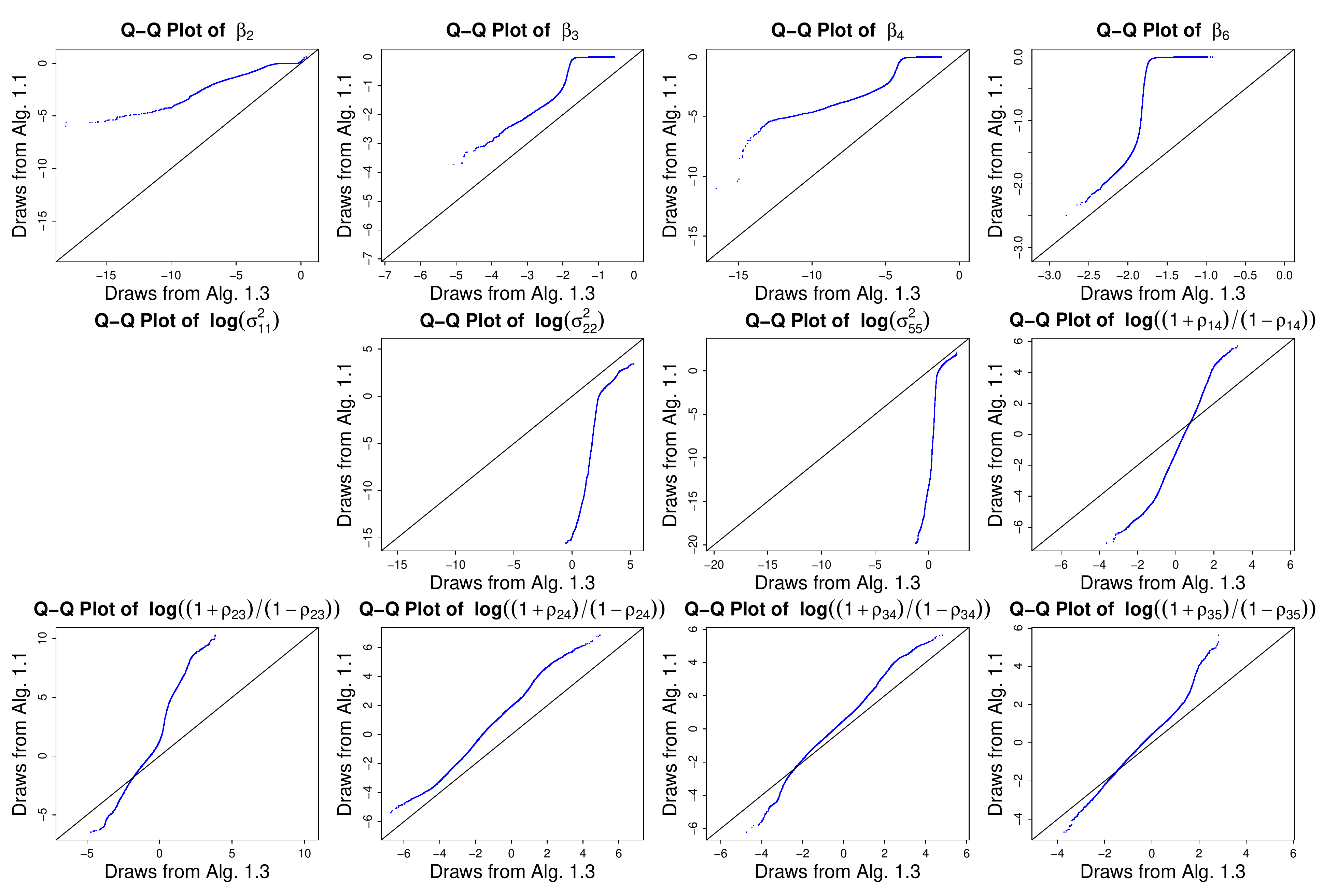}
\end{center}
\caption{Quantile-quantile plots for comparing Algorithms~\ref{alg11} and 1.3 in the margarine-purchase data analysis.}
\label{fig:10}
\end{figure}

\begin{figure}[t]
\begin{center}
\includegraphics[height=4.4in]{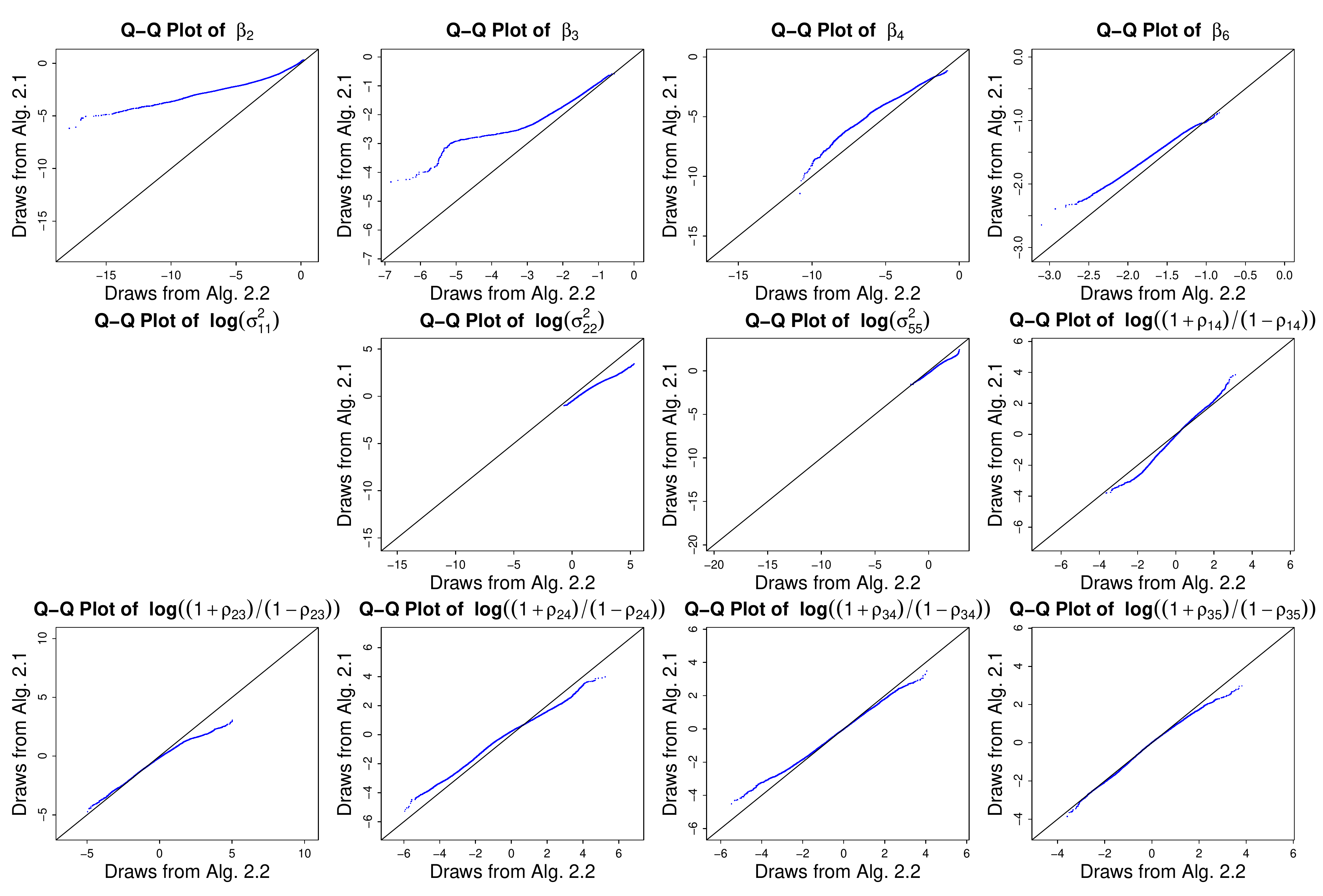}
\end{center}
\caption{Quantile-quantile plots for comparing Algorithms~\ref{alg21} and 2.2 in the margarine-purchase data analysis.}
\label{fig:11}
\end{figure}

\begin{figure}[t]
\begin{center}
\includegraphics[height=4.4in]{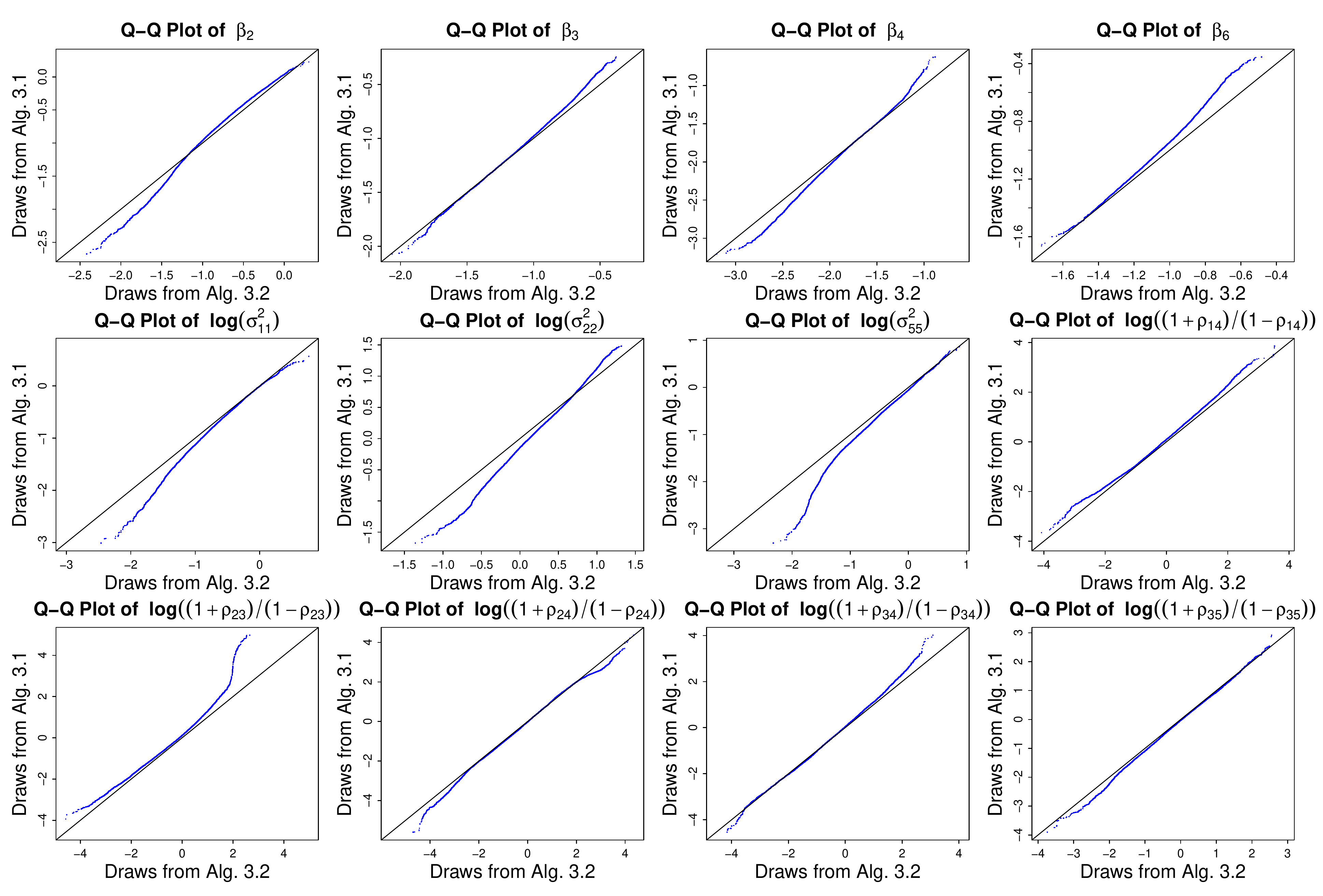}
\end{center}
\caption{Quantile-quantile plots for comparing Algorithms~\ref{alg31} and 3.2 in the margarine-purchase data analysis.}
\label{fig:12}
\end{figure} 

\subsection{Errors in Algorithms and the Corrections}
\label{sec:error}
We refer to Algorithms~1 and~2 of \ivd as Algorithms~\ref{alg11} and~\ref{alg21}. This allows us to clearly number the corrected versions of these algorithms. Similarly, we refer to the algorithm of \bn as Algorithm~\ref{alg31}. Algorithm~\ref{alg11} is displayed here and Algorithms~\ref{alg21} and~\ref{alg31} in Appendices~\ref{app:a} and~\ref{app:b}. 

To obtain posterior samples under the MNP model, Algorithm~\ref{alg11} proceeds by sampling iteratively from $p(\alpha^2,\tilde{W}|Y,\beta,\Sigma)$, $p(\alpha^2,\beta|Y,\tilde{W},\Sigma)$ and $p(\alpha^2,\Sigma|Y,\tilde{W}-\alpha X\beta,\beta)$. The first of these draws is obtained via a sequence of conditional draws, see Step~1(b) of Algorithm~\ref{alg11}. Note that this algorithm marginalizes $\alpha$ out in each step. Algorithm~\ref{alg21} proceeds by sampling iteratively from $p(\alpha^2,\tilde{W}|Y,\beta,\Sigma)$, $p(\alpha^2,\Sigma|Y,\tilde{W}-\alpha X\beta,\beta)$ and $p(\beta|Y,W,\Sigma)$, again using a sequence of conditional draws for updating $\tilde{W}$. Algorithm~\ref{alg21} does not marginalize $\alpha$ out when sampling $\beta$. Algorithm~\ref{alg31} is an adaption of Algorithm~\ref{alg11}. The only difference occurs in Step~3 when sampling $(\alpha^2, \Sigma)$. In Algorithm~\ref{alg11}, $\alpha^2$ is set to the first element of $\tilde{\Sigma}$ in Step~3(b), while it is set to $\tr(\tilde{\Sigma})/p$ in Step~3(b) of Algorithm~\ref{alg31}.

Unfortunately, there are two errors in these algorithms, which may severely alter their stationary distributions, fitted values, and convergence properties. In Algorithm~\ref{alg11}, both errors are in Step~3. The first is rather simple. The transformation from $(Z,\beta^{\nextt},{\alpha}^{\nextt},\tilde{\Sigma}^\star)$ to the original parameterization $(W^{\nextt},\beta^{\nextt},{\alpha}^{\nextt},\Sigma^{\nextt})$ should involve setting
\begin{equation}
W_i^{\nextt}=\left(Z_i+\alpha^{\nextt} X_i\beta^{\nextt}\right)/\alpha^{\nextt},\mbox{ for }i=1,\dots,n,
\label{eq:trans}
\end{equation}
instead of $W_i^{\nextt}=\tilde{W}_i^\star/\alpha^{\nextt}$, see Step~3(b). The correct inverse transformation is necessary to guarantee that the joint stationary distribution of $(W^{\nextt},\beta^{\nextt},{\alpha}^{\nextt},\Sigma^{\nextt})$ is the target posterior distribution. 

The second problem is more subtle. When sampling $\tilde{\Sigma}^\star$ while conditioning on $Y$, $Z$, and $\beta^{\nextt}$, Algorithm~\ref{alg11} uses $\mbox{Inv-Wishart}(n+\nu,\sum_{i=1}^{n}Z_i Z_i^\T)$, see Step~3(a). This however ignores a constraint on $\tilde{\Sigma}^\star$ imposed by $Y$ and the current value of $Z$ and $\beta$. This constraint is on the first diagonal element of $\tilde{\Sigma}^\star$, i.e., ${(\tilde{\sigma}^\star_{11})}^2$. In particular, if we set $\tilde{Z}_i\left(\tilde{\sigma}^\star_{11}\right)=Z_{i}+(\tilde{\sigma}^\star_{11})X_i\beta^{\nextt}$, for $i=1,\dots,n$, the updated value of $\tilde{\sigma}^\star_{11}$ must satisfy
\begin{equation}
\left\{\begin{array}{l l}
  \mbox{max}\left\{\tilde{Z}_{i1}\left(\tilde{\sigma}^\star_{11}\right),\dots,\tilde{Z}_{ip}\left(\tilde{\sigma}^\star_{11}\right)\right\}
<0 & \mbox{ if }Y_i=0\\
  \mbox{max}\left\{0,\tilde{Z}_{i1}\left(\tilde{\sigma}^\star_{11}\right),\dots,\tilde{Z}_{ip}\left(\tilde{\sigma}^\star_{11}\right)\right\}
=\tilde{Z}_{ik}\left(\tilde{\sigma}^\star_{11}\right) & \mbox{ if }Y_i=k
  \end{array}, \right.\mbox{ for }\ i=1,\dots,n.
\label{eq:cons}
\end{equation}
Thus, the conditional distribution of $\tilde{\Sigma}^\star$ given $Y$, $Z$, and $\beta^{\nextt}$ in Step~3(a) should be a constrained $\mbox{inverse-Wishart}$ distribution.

To illustrate the effect of the two corrections to Algorithm~\ref{alg11}, we compare it with two new algorithms:
\begin{description}
\item[\bf Algorithm~1.2:] This is a partial correction to Algorithm~\ref{alg11}. The only difference is that Algorithm~1.2 transforms $(Z,\beta^{\nextt},{\alpha}^{\nextt},\tilde{\Sigma}^\star)$ to $(W^{\nextt},\beta^{\nextt},{\alpha}^{\nextt},\Sigma^{\nextt})$ using~(\ref{eq:trans}) in Step~3(b). 

\item[\bf Algorithm~1.3:] This algorithm completely corrects Algorithm~\ref{alg11}. In particular, Steps~0, 1, and 2 of Algorithm~1.3 are the same as Algorithm~\ref{alg11}. In Step~3(a), however, Algorithm~1.3 updates $\tilde{\Sigma}^\star$ by sampling from a constrained $\mbox{inverse-Wishart}$ distribution, that is, 

$$\tilde{\Sigma}^\star\sim\mbox{Inv-Wishart}\left(n+\nu,\sum\limits_{i=1}^{n}Z_i Z_i^\T\right)
\mbox{ subject to the constraint in~(\ref{eq:cons})}.$$
This is accomplished by simple rejection sampling; we iteratively sample from the unconstrained inverse-Wishart distribution until~(\ref{eq:cons}) is satisfied. Finally, in Step~3(b), Algorithm~1.3 sets ${\alpha}^{\nextt}=\tilde{\sigma}_{11}^\star$, $\Sigma^{\nextt}=\tilde{\Sigma}^\star/{\left(\alpha^{\nextt}\right)}^2$, and $W_i^{\nextt}=(Z_i+\alpha^{\nextt} X_i\beta^{\nextt})/\alpha^{\nextt}$.
\end{description}

Algorithms~\ref{alg21} and~\ref{alg31} are adaptions of Algorithm~\ref{alg11}. Thus, both corrections affect these algorithms as well. The corrected versions of Algorithms~\ref{alg21} and~\ref{alg31} are called Algorithms~2.2 and~3.2 respectively. See Appendices~\ref{app:a} and~\ref{app:b} for details of these algorithms. 

\section{Simulation Study}
\label{sec:simu}
We use a simulation study to illustrate the differences in the convergence properties of Algorithms~\ref{alg11},~1.2, and~1.3, Algorithms~\ref{alg21} and~2.2, and Algorithms~\ref{alg31} and~3.2. We set $n=50$, $p=2$, $q=2$, $\beta=(-\sqrt{2},1)$, $\Sigma=\scalefont{0.75}{\left(\begin{array}{cc}
1 & 0.5\\
0.5 & 1\end{array}\right)}$. For $X_i=\scalefont{0.75}{\left(\begin{array}{cc}
X_{i1,1} & X_{i1,2}\\
X_{i2,1} & X_{i2,2}\end{array}\right)}$, we sample $X_{ij,1}$ ($j=1,2$) from a uniform distribution on $(-0.5,0.5)$ for $i=1,\dots,25$, on $(0.4,1.5)$ for $i=26,\dots,50$, and sample $X_{ij,2}$ ($j=1,2$) from a uniform distribution on $(-1,1)$ for $i=1,\dots,25$, on $(0.8,3)$ for $i=26,\dots,50$. We specify the prior distribution of $\Sigma$ and $\alpha^2$ as in Section~\ref{sec:model}, with $\nu=p$, $\alpha_0^2=\nu$, and $S={\rm Diag}(1,1)$, and for $\beta$, as $\beta\sim\N_q[0,{\rm Diag}(100,100)]$. For each algorithm, we run a chain of length 15,000 and discard the first 5,000 draws.

Figure~\ref{fig:1} presents the posterior samples of $W_{7}$, $W_{8}$, and $W_{34}$ obtained with Algorithms~1.2 and~1.3 respectively. The draws obtained with Algorithm~1.2 that do not adhere to the constraint~(\ref{eq:cons}) are colored in red, which illustrates the second problem of Algorithm~\ref{alg11} described in Section~\ref{sec:error}. Such draws are rejected in Step~3(a) of Algorithm~1.3.

Most importantly, Algorithms~\ref{alg11} (or 1.2),~\ref{alg21}, and~\ref{alg31} do not return draws from the target posterior distribution. Figure~\ref{fig:9} shows the quantile-quantile plots of parameters to compare the stationary distributions of original and corrected algorithms. The first row of Figure~\ref{fig:9} compares Algorithms~\ref{alg11} and 1.3. The distributions of $\beta$ differ slightly for the two algorithms, while the distributions of $\Sigma$ differ significantly, especially the correlation parameter, $\rho_{12}=\sigma_{12}/(\sigma_{11}\sigma_{22})$. For Algorithms~1.2 and~1.3 (not shown), the distributions of $\beta$ are again similar, while the distributions of $\Sigma$ again differ, but not as severely as Algorithms~\ref{alg11} and~1.3. The second row shows the quantile-quantile plots that compare Algorithms~\ref{alg21} and~2.2. The distributions of both $\beta$ and $\Sigma$ are slightly different for the two algorithms. The last row of Figure~\ref{fig:9} compares Algorithms~\ref{alg31} and 3.2. The distributions of $\beta$ are rather similar for the two algorithms, while the distributions of $\Sigma$ are different, particularly $\rho_{12}$ and $\sigma^2_{22}$. 

Figures~\ref{fig:2} and~\ref{fig:4} show the sampling results of Algorithms~\ref{alg11} and 1.3 respectively. The columns in both figures correspond to trace plots, autocorrelation plots, and histograms. The rows correspond to four parameters, namely, $\beta_1$, $\beta_2$, $\logr\left(\frac{1+\rho_{12}}{1-\rho_{12}}\right)$, and $\logr(\sigma^2_{22})$. Algorithm~1.3 has better convergence properties than Algorithm~\ref{alg11} for all the four parameters in terms of mixing and autocorrelation. The convergence of Algorithm~1.2 is better than Algorithm~\ref{alg11}, but not as good as Algorithm~1.3. Moreover, Algorithms~2.2 and 3.2 have slightly better convergence properties than Algorithms~\ref{alg21} and~\ref{alg31} respectively. We omit the corresponding plots for Algorithms~1.2,~\ref{alg21}, 2.2,~\ref{alg31}, and~3.2 to save space.

To further compare the convergence properties of these algorithms, we compute the {\it effective sample size} (ESS), defined by
\begin{equation}
{\rm ESS}(\theta)=\frac{T}{1+2\sum_{t=1}^\infty \rho_{t}(\theta)},
\label{eq:ess}
\end{equation}
where $T$ is the total posterior sample size and $\rho_{t}(\theta)$ is the lag-$t$ autocorrelation of the parameter $\theta$. ESS gives an estimate of the equivalent number of independent iterations that a Markov chain represents, and it indicates how well the chain mixes, see~\citet{kass:etal:98} and \citet{liu:01:book}. We use the function ``effectiveSize'' in the R package \emph{coda} to calculate ESS. To account for the required CPU time, we compare ESS per second of these algorithms. The larger the value, the more efficient is the algorithm. The first three columns in Table~\ref{table:ess} present the ESS per second for each parameter for Algorithms~\ref{alg11}--1.3, respectively. The fourth and fifth columns correspond to Algorithms~\ref{alg21} and~2.2, and the last two columns correspond to Algorithms~\ref{alg31} and~3.2. We find that in terms of ESS per second, Algorithm~1.3 is more efficient than Algorithm~\ref{alg11} even though it is more computationally demanding per iteration, and it performs similarly to Algorithm~1.2. Algorithm~2.2 performs similarly to Algorithm~\ref{alg21}, and Algorithm~3.2 performs slightly better than Algorithm~\ref{alg31}. 

\section{Data Analysis}
\label{sec:real}
For a further comparison of the algorithms, we consider a data set describing margarine purchases which is available in the \emph{bayesm} package of R. Following BN, we limit analysis to purchases of six brands: ``Parkay stick'', ``Blue Bonnet stick'', ``Fleischmanns stick'', ``House brand stick'', ``Generic stick'', and ``Shedd Spread tub'', and only consider the first purchase
of one of these brands for each household. This results in a dataset consisting of $n=507$ observations. 

We set ``Parkay stick'' as the base category, and $p=5$. Again following BN, we set up a model that only includes intercept terms for the other five categories and a coefficient for log prices. Thus $q=6$, and $X_i=[I_p, g_i]$, where $I_p$ is the identity matrix and $g_i$ is the $p$-vector of differences in log prices between each category and the base. We again specify the prior distribution for $\Sigma$ and $\alpha^2$ as in Section~\ref{sec:model}, with $\nu=p$, $\alpha_0^2=\nu$, and $S={\rm Diag}(1,\dots,1)$, and for $\beta$, as $\beta\sim\N_q[0,{\rm Diag}(100,\dots,100)]$. When implementing Algorithms~\ref{alg11},~1.3,~\ref{alg21} and~2.2, we set the variance corresponding to ``Blue Bonnet stick'' as one. For each algorithm, we run a chain of length 300,000, discard the first 100,000 draws, and thin the rest draws by 10. In this way we obtain 20,000 draws from each algorithm. 

Figures~\ref{fig:10},~\ref{fig:11}, and~\ref{fig:12} present the quantile-quantile plots of selected parameters correspondingly sampled with Algorithms~\ref{alg11} and~1.3, Algorithms~\ref{alg21} and~2.2, and Algorithms~\ref{alg31} and~3.2 respectively. The parameters we consider are $\beta_2$, $\beta_3$, $\beta_4$, $\beta_6$, $\logr(\sigma^2_{11})$, $\logr(\sigma^2_{22})$, $\logr(\sigma^2_{55})$, $\logr\left(\frac{1+\rho_{14}}{1-\rho_{14}}\right)$, $\logr\left(\frac{1+\rho_{23}}{1-\rho_{23}}\right)$, $\logr\left(\frac{1+\rho_{24}}{1-\rho_{24}}\right)$, $\logr\left(\frac{1+\rho_{34}}{1-\rho_{34}}\right)$, and $\logr\left(\frac{1+\rho_{35}}{1-\rho_{35}}\right)$. They are selected because their stationary distributions show relatively obvious difference for all three pairs of original and corrected algorithms. We find that Algorithms~\ref{alg11},~\ref{alg21}, and~\ref{alg31} all fail to deliver draws from the target posterior distribution. The situation is most substantial for Algorithm~\ref{alg11}. Moreover, in terms of autocorrelation, Algorithm~1.3 performs substantially better than Algorithm~\ref{alg11}, while Algorithms~2.2 and~3.2 perform similarly as Algorithms~\ref{alg21} and~\ref{alg31} respectively. In addition, Algorithms~1.3,~2.2, and~3.2 take around 15\% more computational time than Algorithms~\ref{alg11},~\ref{alg21}, and~\ref{alg31} respectively.

\section{Conclusion}
\label{sec:con}
The algorithms of \ivd and \bn are implemented in the popular R package \emph{MNP} and are widely used for fitting MNP models. We point out errors in these algorithms and propose corrections. Using both a simulation study and a real-data analysis, we illustrate the difference between the original and corrected algorithms. From these analyses, we find that the errors can significantly affect the final results, especially in that they alter the stationary distribution and hence the fitted parameters. Considering the popularity of these algorithms, it is important that they are corrected. We have done so here and will do it soon in the \emph{MNP} package. The corrected algorithms require some what more computational time due to the additional rejection sampling steps, however, the extra computational time is small and at least in some cases it is made up by the improved autocorrelation of the corrected algorithms.

\bibliographystyle{natbib}
\bibliography{my}

\clearpage
\appendix

\counterwithin{figure}{section}

\bigskip
\begin{center}
{\Large\bf APPENDIX: Details of Algorithms~\ref{alg21}--3.2}
\end{center}

\section{Algorithms~\ref{alg21} and 2.2}
\label{app:a}

\begin{vartalgorithm}{2.1}
\caption{}\label{alg21}
\begin{algorithmic}
\footnotesize{\State {\bf Step~0:} Initialize parameters $t=0$, $\beta^{\ini}$, $\alpha^{\ini}$, $\Sigma^{\ini}$ and $W^{\ini}$.
\smallskip
\While {$t<T$}
\smallskip
\State {\bf Step~1:} Update $\left({(\alpha^2)}^{\star},Z\right)$ from $p(\alpha^2,Z|Y,\beta^{\cutt}, \Sigma^{\cutt})$ by

\State (a) sampling ${(\alpha^2)}^{\star}$ from $p(\alpha^2|\Sigma^{\cutt})$: ${(\alpha^2)}^{\star}\sim \alpha_0^2\tr\left(S{\Sigma^{\cutt}}^{-1}\right)/\chi^2_{\nu p}$;

\State (b) sampling $Z$ from $p(Z|Y,{(\alpha^2)}^{\star},\beta^{\cutt},\Sigma^{\cutt})$: 
\For{$i:=1,\dots,n$}
\For{$k:=1,\dots,p$}

\State sampling $W_{ik}^\star$ via $p(W_{ik}|Y_i,W_{i,-k}^\star,\beta^{\cutt}, \Sigma^{\cutt})$: $W_{ik}^\star\sim {\rm TN}(\mu_{ik},\tau_{ik}^2)$, see Appendix~\ref{app:c} for details;
\EndFor

\State Set $Z_i=\alpha^{\star}({W}_i^\star-X_i\beta^{\cutt})$.
\EndFor

\smallskip
\State {\bf Step~2:} Update $\left({(\alpha^2)}^{\nextt},\Sigma^{\nextt}\right)$ via $p(\alpha^2,\Sigma|Y,Z, \beta^{\cutt})$ by

(a) sampling $\tilde{\Sigma}^\star$ from $p(\tilde{\Sigma}|Y,Z,\beta^{\cutt})$:
$$\tilde{\Sigma}^\star\sim \mbox{Inv-Wishart}\left[n+\nu,\sum_{i=1}^{n}Z_i {Z_i}^\T\right];$$

(b) setting ${\alpha}^{\nextt}=\tilde{\sigma}_{11}^\star$, $\Sigma^{\nextt}=\tilde{\Sigma}^\star/{\left(\alpha^{\nextt}\right)}^2$, and $W_i^{\nextt}=(Z_i+\alpha^{\nextt} X_i\beta^{\cutt})/\alpha^{\nextt}$.

\smallskip
\State {\bf Step~3:} Update $\beta^{\nextt}$ via $p(\beta|Y,W^{\nextt}, \Sigma^{\nextt})$:
$$\beta^{\nextt}\sim \N_q\left[\hat{\beta},
{\left(\sum_{i=1}^n X_i^\T {\Sigma^{\nextt}}^{-1}X_i +A^{-1}\right)}^{-1}\right],$$
\State where $\hat{\beta}={\left(\sum_{i=1}^n X_i^\T {\Sigma^{\nextt}}^{-1}X_i +A^{-1}\right)}^{-1}
\left(\sum_{i=1}^n X_i^\T {\Sigma^{\nextt }}^{-1}W_i^{\nextt}\right)$.

\medskip
\State \Return $\beta^{\nextt}$, $\Sigma^{\nextt}$, and $W^{\nextt}$
\State $t+1 \gets t$
\EndWhile}
\end{algorithmic}
\end{vartalgorithm}

Algorithm~2 of \ivd does not marginalize $\alpha$ out when updating $\beta$. We call this algorithm Algorithm~\ref{alg21} in this paper. Algorithm~\ref{alg21} can be used when the prior mean of $\beta$, $\beta_0$, is not equal to zero, while Algorithm~\ref{alg11} can not. 

The error arises in Step~2(a), which is the same as the error in Step~3(a) of Algorithm~\ref{alg11}. Thus the correction to Algorithm~\ref{alg21} is 
\begin{description}
\item[\bf Algorithm~2.2:] Steps~0, 1, and 3 of Algorithm~2.2 are the same as Algorithm~\ref{alg21}. In Step~2(a), however, Algorithm~2.2 updates $\tilde{\Sigma}^\star$ by sampling from a constrained $\mbox{inverse-Wishart}$ distribution, that is, 

$$\tilde{\Sigma}^\star\sim\mbox{Inv-Wishart}\left(n+\nu,\sum\limits_{i=1}^{n}Z_i Z_i^\T\right)
\mbox{ subject to the constraint in~(\ref{eq:cons})}.$$
Note that $\beta^{\nextt}$ in $\tilde{Z}_i\left(\tilde{\sigma}^\star_{11}\right)$ of the constraint~(\ref{eq:cons}) should be replaced by $\beta^{\cutt}$ in Algorithm~2.2. 
\end{description}

\section{Algorithms~\ref{alg31} and 3.2}
\label{app:b}

\begin{vartalgorithm}{3.1}
\caption{}\label{alg31}
\begin{algorithmic}
\footnotesize{\State {\bf Step~0:} Initialize parameters $t=0$, $\beta^{\ini}$, $\alpha^{\ini}$, $\Sigma^{\ini}$ and $W^{\ini}$.
\smallskip
\While {$t<T$}
\smallskip
\State {\bf Step~1:} Update $\left({(\alpha^2)}^{\star},\tilde{W}^\star\right)$ via $p(\alpha^2,\tilde{W}|Y,\beta^{\cutt}, \Sigma^{\cutt})$ by

\State (a) sampling ${(\alpha^2)}^{\star}$ from $p(\alpha^2|\Sigma^{\cutt})$: ${(\alpha^2)}^{\star}\sim \alpha_0^2\tr\left(S{\Sigma^{\cutt}}^{-1}\right)/\chi^2_{\nu p}$;

\State (b) sampling $\tilde{W}^\star$ from $p(\tilde{W}^\star|Y,{(\alpha^2)}^{\star},\beta^{\cutt},\Sigma^{\cutt})$: 
\For{$i:=1,\dots,n$}
\For{$k:=1,\dots,p$}

\State sampling $W_{ik}^\star$ from $p(W_{ik}|Y_i,W_{i,-k}^\star,\beta^{\cutt}, \Sigma^{\cutt})$: $W_{ik}^\star\sim {\rm TN}(\mu_{ik},\tau_{ik}^2)$, see Appendix~\ref{app:c} for details;
\EndFor

\State Set $\tilde{W}_i^\star=\alpha^{\star}{W}_i^\star$. 
\EndFor

\smallskip
\State {\bf Step~2:} Update $\left({(\alpha^2)}^{\star},\beta^{\nextt}\right)$ via $p(\alpha^2,\beta|Y,\tilde{W}^\star, \Sigma^{\cutt})$ by 

(a) sampling ${(\alpha^2)}^{\star}$ from $p(\alpha^2|Y,\tilde{W}^\star, \Sigma^{\cutt})$:
$${(\alpha^2)}^{\star}\sim
\frac{\sum_{i=1}^n{(\tilde{W}_i^\star-X_i\hat{\beta})}^\T{\Sigma^{\cutt}}^{-1}(\tilde{W}_i^\star-X_i\hat{\beta})
+\hat{\beta}^\T A^{-1}\hat{\beta}+\tr\left(\tilde{S}{\Sigma^{\cutt }}^{-1}\right)}{\chi^2_{(n+\nu) p}},$$
\State where $\hat{\beta}={\left(\sum_{i=1}^n X_i^\T {\Sigma^{\cutt }}^{-1}X_i +A^{-1}\right)}^{-1}
\left(\sum_{i=1}^n X_i^\T {\Sigma^{\cutt }}^{-1}\tilde{W}_i^\star\right)$;

(b) sampling $\tilde{\beta}^\star$ from $p(\tilde{\beta}|Y,\tilde{W}^\star, {(\alpha^2)}^\star,\Sigma^{\cutt})$:
$$\tilde{\beta}^\star\sim \N_q\left[\hat{\beta},
{(\alpha^2)}^{\star}{\left(\sum_{i=1}^n X_i^\T {\Sigma^{\cutt }}^{-1}X_i +A^{-1}\right)}^{-1}\right],$$
\State and setting $\beta^{\nextt}=\tilde{\beta}^\star/\alpha^{\star}$.

\smallskip
\State {\bf Step~3:} Update $\left({(\alpha^2)}^{\nextt},\Sigma^{\nextt}\right)$ via $p(\alpha^2,\Sigma|Y,\tilde{W}^\star, \beta^{\nextt})$ by 

(a) sampling $\tilde{\Sigma}^\star$ from $p(\tilde{\Sigma}|Y,Z,\beta^{\nextt})$: 
$$\tilde{\Sigma}^\star\sim \mbox{Inv-Wishart}\left(n+\nu,\sum_{i=1}^{n}Z_i Z_i^\T\right),$$
\State where $Z_i=\tilde{W}_i^\star-\alpha^\star X_i \beta^{\nextt}$; 

(b) setting ${\alpha}^{\nextt}=\sqrt{\tr(\tilde{\Sigma}^\star/p)}$, $\Sigma^{\nextt}=\tilde{\Sigma}^\star/{\left(\alpha^{\nextt}\right)}^2$, $\beta^{\nextt}=\tilde{\beta}^\star/\alpha^{\nextt}$, and $W^{\nextt}=\tilde{W}^\star/\alpha^{\nextt}$.
\medskip
\State \Return $\beta^{\nextt}$, $\Sigma^{\nextt}$, and $W^{\nextt}$
\State $t+1 \gets t$
\EndWhile}
\end{algorithmic}
\end{vartalgorithm}

We call the algorithm of \bn Algorithm~\ref{alg31} in this paper. Algorithm~\ref{alg31} is almost the same as Algorithm~\ref{alg11}. The only difference is Step~3(b). Specifically, first, in Algorithm~\ref{alg31}, $\alpha^2$ in this step is set to $\tr(\tilde{\Sigma})/p$, while in Algorithm~\ref{alg11}, $\alpha^2$ is set to the first element of $\tilde{\Sigma}$; second, Algorithm~\ref{alg31} sets $\beta=\tilde{\beta}/\alpha$ in Step~3(b), while Algorithm~\ref{alg11} not.

Besides applying the two corrections to Step~3 of Algorithm~\ref{alg31}, we further remove ``$\beta^{\nextt}=\tilde{\beta}^\star/\alpha^{\nextt}$'' in Step~3(b) of Algorithm~\ref{alg31}, because we update $\tilde{\Sigma}^\star$ conditioning on $(Y, Z, \beta^{\nextt})$, not on $(Y,\tilde{W}^\star, \tilde{\beta}^{\star})$. Thus, we get 
\begin{description}
\item[\bf Algorithm~3.2:] This algorithm completely corrects Algorithm~\ref{alg31}. In particular, Steps~0, 1, and 2 of Algorithm~3.2 are the same as Algorithm~\ref{alg31}. In Step~3(a), however, Algorithm~3.2 updates $\tilde{\Sigma}^\star$ by sampling from a constrained $\mbox{inverse-Wishart}$ distribution, that is, 

$$\tilde{\Sigma}^\star\sim\mbox{Inv-Wishart}\left(n+\nu,\sum\limits_{i=1}^{n}Z_i Z_i^\T\right)
\mbox{ subject to the constraint ${(10^\star)}$}.$$
Constraint $(10^\star)$ is an adaption of (\ref{eq:cons}) by replacing $\tilde{\sigma}^\star_{11}$ with $r=\sqrt{\tr(\tilde{\Sigma}^\star/p)}$. Specifically, $\tilde{Z}_i(r)=Z_{i}+r X_i\beta^{\nextt}$, for $i=1,\dots,n$. The updated value of $r$ must satisfy
\begin{equation}
\left\{\begin{array}{l l}
  \mbox{max}\left\{\tilde{Z}_{i1}(r),\dots,\tilde{Z}_{ip}(r)\right\}
<0 & \mbox{ if }Y_i=0\\
  \mbox{max}\left\{0,\tilde{Z}_{i1}(r),\dots,\tilde{Z}_{ip}(r)\right\}
=\tilde{Z}_{ik}(r) & \mbox{ if }Y_i=k
  \end{array}, \right.\mbox{ for }\ i=1,\dots,n.\tag{$10^\star$}
\label{eq:cons2}
\end{equation}
Finally, in Step~3(b), Algorithm~3.2 sets ${\alpha}^{\nextt}=\sqrt{\tr(\tilde{\Sigma}^\star/p)}$, $\Sigma^{\nextt}=\tilde{\Sigma}^\star/{\left(\alpha^{\nextt}\right)}^2$, and $W_i^{\nextt}=(Z_i+\alpha^{\nextt} X_i\beta^{\nextt})/\alpha^{\nextt}$.
\end{description}

\section{Details of Sampling $W$ in Step 1(b) of Algorithms~\ref{alg11}--3.2}
\label{app:c}
Updating $W$ in Step~1(b) of Algorithms~\ref{alg11}--3.2 consists of sampling from a series of univariate truncated normal distributions, that is, for $i=1,\dots,n$ and $k=1,\dots,p$, 
$$W_{ik}^\star\sim {\rm TN}(\mu_{ik},\tau_{ik}^2),$$ 
where $\mu_{ik}=X_{ik}\beta^{\cutt}+\Sigma^{\cutt}_{k,-k}\Sigma^{\cutt -1}_{-k,-k}(W_{i,-k}^\star-X_{i,-k}\beta^{\cutt})$ with ${W}_{i,-k}^\star=\left(W_{i1}^\star,\dots,W_{i,(k-1)}^\star,W_{i,(k+1)}^{\cutt},\dots,W_{ip}^{\cutt}\right)$, and $\tau^2_{ik}={\left(\sigma^{\cutt}_{kk}\right)}^2-\Sigma^{\cutt}_{k,-k}\Sigma^{\cutt -1}_{-k,-k}\Sigma^{\cutt}_{-k,k}$; $X_{ik}$ is the $k\order$ row of $X_i$, and $X_{i,-k}$ is the sub-matrix of $X_i$ with $X_{ik}$ removed. The constraint on $W_{ik}^\star$ is, $W_{ik}^\star\ge{\rm max}\{0,W_{i,-k}^\star\}$, if $Y_{i}=k$; $W_{ik}^\star<0$, if $Y_{i}=0$; and $W_{ik}^\star\le {\rm max}\{0,W_{ij}^\star$\}, if $Y_{i}=j\neq k$. 

If the constraint on $W_{ik}^\star$ has the form, $W_{ik}^\star\ge w$, and $w\le 0$, we update $W_{ik}^\star$ with simple rejection sampling: we iteratively sample from the unconstrained normal distribution until $W_{ik}^\star\ge w$ is satisfied. If $W_{ik}^\star\ge w$, but $w> 0$, we update $W_{ik}^\star$ with the exponential rejection sampling proposed by \citet{robert:95}. If the constraint on $W_{ik}^\star$ has the form $W_{ik}^\star\le w$, we can apply the above sampling scheme with slight adaption, since $-W_{ik}^\star\ge -w$.
\end{document}